\documentclass[twocolumn,showpacs]{revtex4}
\usepackage{graphics,epsf,bm}
\begin{document}
\title{Comparative study of hyperon-nucleon interactions of quark model
and chiral effective field theory by low-momentum equivalent interactions
and $G$ matrices}
\author{M. Kohno}
\affiliation{Physics Division, Kyushu Dental College,
Kitakyushu 803-8580, Japan}

\begin{abstract}
Hyperon-nucleons interactions constructed by two frameworks, the Kyoto-Niigata
SU$_6$ quark model and the chiral effective field theory, are compared by
investigating equivalent interactions in a low-momentum space and in addition
by calculating hyperon single-particle potentials in the lowest-order Brueckner
theory in symmetric nuclear matter. Two descriptions are shown to give similar
matrix elements in most channels after renormalizing high momentum components.
Although the range of the $\Lambda N$ interaction is different in two
potentials, the $\Lambda$ single-particle potential in nuclear matter is very
similar. The $\Sigma$-nucleus and $\Xi$-nucleus potentials are also found to
be similar. These predictions are to be confronted with forthcoming
experimental data.
\end{abstract}

\maketitle

\section{Introduction}
It is basically important to obtain a realistic potential description of
baryon-baryon interactions for understanding the properties of baryons and
baryonic systems. Baryon-baryon interactions in the strangeness $S=-1$ and
$S=-2$ sectors have not been well regulated by experiments, except for
a fair amount of data for $\Lambda$ hypernuclei. The construction of these
potentials has to rely on an underlying theoretical framework, such as a one
boson-exchange potential (OBEP) picture, a constituent quark model, and
a chiral effective field theory (EFT). Predictions of these different
potential descriptions for hypernuclear phenomenon, for example $\Sigma$
and $\Xi$ hypernuclear bound states, multi hyperon systems, and properties
of neutron star matter, naturally vary. Future experimental data will
constrain the parameters to allow more solid predictions. Before the
experiment, however, it is interesting and important to make a comparison
between presently available potential parametrizations to elucidate the
character of the underlying theoretical frameworks.

As is known in the nucleon-nucleon ($NN$) interaction, the direct comparison
of the bare potential is not meaningful. We have to consider some effective
interactions and quantities closely connected to experimental observables,
such as s.p. potentials in the nuclear medium. In this context, equivalent
interactions in a low-momentum space \cite{BOG} have become a useful
tool to figure out the properties of baryon-baryon interactions without
being obscured by uncertainties in the
description of the short-range part. We call an effective interaction in a
restricted space which reproduces the same eigenvalues or $T$-matrices
in that space as those of the original full-space interaction an equivalent
interaction.

In Ref. \cite{KOKF}, we reported the comparison of low-momentum space
equivalent interactions of the Nijmegen OBEP nsc97f \cite{NSC} and the
Kyoto-Niigata SU$_6$ quark-model potential fss2 \cite{FSN} for $\Lambda N$
and $\Sigma N$ interactions, and showed the property of the $\Xi N$
interaction of fss2. For the $\Lambda N$ case, two models provide very
similar matrix elements
in a low-momentum space, although the short-range part
is considerably different. On the contrary, there is difference in
the $\Sigma N$ interaction.

In this paper we extend the study to consider the potential by the
chiral EFT \cite{CEFT1,CEFT2} and make a comparison
with the quark-model potential fss2 in two ways; namely by investigating
equivalent interactions in a low-momentum space and hyperon
s.p. potentials in nuclear matter in the framework of the lowest-order
Brueckner theory. The elimination of the high-momentum components
by considering low-momentum space equivalent interaction
enables us to concentrate on features of the $YN$ interaction
relevant to low-energy experimental hypernuclear observables.
To consider the implication of baryon-baryon interactions to experimental
quantities, it is not sufficient to study the low-momentum interaction.
Important correlations inside a low-momentum space and many-body
correlations in a high-momentum including the components in a
high-momentum space have to be incorporated to obtain physically
meaningful quantities. The standard way in nuclear physics is the
Brueckner theory. It deals with singular short-range parts of the
baryon-baryon interaction, and at the same time incorporates
important many-body effects through the Pauli principle and the
dispersion effects. Thus we calculate hyperon s.p. potentials in
symmetric nuclear matter in the Brueckner theory. The feasible
lowest-order calculation accounts for semi-quantitatively the
structure dependence of the hyperon-nucleon interactions in the
nuclear medium. The s.p. potential is one of most important
quantities connected with baryon properties in the nuclear medium,
although they are not direct observables. Therefore the hyperon
s.p. potentials in the LOBT in symmetric nuclear matter provide
a further insight into the properties of the bare hyperon-nucleon
interactions.

The fss2 potential is the most recent model by the Kyoto-Niigata
group \cite{FSN,FU96a}, in which an effective gluonic interaction and
long-ranged one-boson exchanges between quarks are considered in the
resonating group method (RGM) for two constituent-quark clusters.
This fss2 \cite{FSN} achieves comparable accuracy in the $NN$ sector to
modern realistic $NN$ potentials. The extension of the potential to the
strangeness $S=-1$ and $S=-2$ sectors on the basis of the parameters fixed
in the $NN$ sector has been shown \cite{FSN} to be less ambiguous than
the OBEP parametrization. In fact, the prediction of the overall
repulsive nature of the $\Sigma$-nucleus potential before experiments
is supported by analyses \cite{NOUM,HH,MK} of the $(\pi^-,K^+)$ $\Sigma$
production inclusive spectra \cite{NOUM,SAHA}. The microscopic
calculation of the $\Sigma$-nucleus s.p. potential in finite nuclei
\cite{KF09} further demonstrated that the fss2 potential actually
reproduces the subtle structure of the weak surface attraction and
the repulsion inside a nucleus which is indicated by the analyses
\cite{BFG94} of the shift and width of $\Sigma^-$ atomic levels.

The chiral EFT potentials in the strangeness $S=-1$ and $S=-2$ sector
have been recently developed by the J\"{u}lich group \cite{CEFT1,CEFT2},
as the extension of the nucleon-nucleon case \cite{EGM}. This description
uses pseudoscalar-meson exchanges and flavor SU$_3$ invariant
contact terms, regularized by a cut-off mass of around 600 MeV.
At present the interaction is derived in the leading order.
Parameters of the contact terms, 5 in number in the $S=-1$ sector
and an additional one parameter in the $S=-2$ sector, are determined
by fitting to available experimental data. Because the description for
the short-range part is considerably different from that of fss2,
it is worthwhile to compare two potentials.

In Sec. II, we briefly describe the basics of the equivalent interaction
theory in a model space. Results of numerical calculations in $^1S_0$
and $^3S_1$ channels are presented in Sec. III for $\Lambda N$, $\Sigma N$,
and $\Xi N$ interactions. We also present the $\Lambda$, $\Sigma$,
and $\Xi$ s.p. potentials in symmetric nuclear matter at various Fermi
momenta between $k_F=0.75$ and $1.45$ fm$^{-1}$. Section IV summarizes
the results of the present paper.

\section{Equivalent interaction}
Suzuki and Lee \cite{SL80,LS80} proposed in 1980 the basic idea to
construct the energy-independent hermitian equivalent Hamiltonian
in a model space $P$. Their consideration is closely related to the
recent development of low-momentum interactions \cite{BOG}. It is
elementary to observe that the eigenvalues of the original Hamiltonian
$H$ do not change when $H$ is transformed by a similarity transformation,
namely by a regular matrix $X$ and its inverse $X^{-1}$ as
$H\Rightarrow H' \equiv X^{-1}HX$. It is easy to see that if a
decoupling condition $QX^{-1}HXP=0$ holds with $Q=1-P$, $PX^{-1}HXP$
becomes the equivalent Hamiltonian $H_{eff}$ in the model space $P$.
Thus the task to find $H_{eff}$ is reduced to determine $X$ which
satisfies $QX^{-1}HXP=0$. It is sufficient first to consider a regular
matrix $X$ in the following form.
\begin{equation} X=\left( \begin{array}{cc}
            1, 0 \\
            \omega, 1
        \end{array} \right), \hspace{1em}\mbox{then}\hspace{1em}
        X^{-1}=\left( \begin{array}{cc}
            1, 0 \\
            -\omega, 1
        \end{array} \right).
\end{equation}
The mapping matrix $\omega = Q\omega P$, which connects the $P$ and $Q$
spaces, plays a central role in the construction of $H_{eff}$.
The decoupling condition $QX^{-1}HXP=0$ now reads:
\begin{equation}
 QHP+QHQ\omega -\omega PHP-\omega PHQ\omega =0.
\end{equation}
Because this is a non-linear equation for $\omega$, we have to use some
iteration method to solve it. Determining the mapping operator $\omega$,
we obtain an energy-independent equivalent Hamiltonian in the model
space $P$ as $H_{eff}= PHP +PHQ\omega P$. This equivalent Hamiltonian
is not hermitian at this stage. If we utilize a unitary matrix $\tilde{X}$
in the following Okubo form \cite{OKU} constructed from $\omega$ of
Eq. (1) to transform the original $H$, we obtain the hermitian Hamiltonian.
\begin{equation}\tilde{X} = \left( \begin{array}{cc}
            1, -\omega^\dagger \\
            \omega, \hspace{0.5em}1
        \end{array} \right) \left(\begin{array}{cc}
            1+\omega^\dagger \omega ,0\hspace*{1.5em}\\
            \hspace*{1.5em}0, 1+\omega \omega^\dagger
        \end{array}  \right)^{-1/2},
\end{equation}
Subtracting the kinetic part, we can define an equivalent interaction
in the model space. In the case of the equivalent interaction in
a two-body problem, for example the elimination of high-momentum
components, the procedure is transparent, because many-body
correlations do not appear.

The actual calculation of the mapping operator $\omega$ is carried out by
the method-2 in Ref. \cite{FUJI}. The extension to the hyperon-nucleon case,
in which several baryon-channels couple each other and there appears an
antisymmetric spin-orbit coupling absent in the $NN$ interaction, is
straightforward. However, we encounter numerical troubles in some cases,
i.e. in the $T=\frac{1}{2}$ $\Sigma N$ $^1S_0$ and $^3S_1$ channels and
the $T=0$ $\Xi N$ $^1S_0$ channel, when the threshold of another baryon
channel is located in the low-momentum space. The extended method-2 yields
oscillatory behavior of the matrix elements of the equivalent interaction
as a function of the momentum that varies as mesh points are altered.
One tentative remedy is to use rather coarse mesh points to obtain
smooth $k$-dependence. But, this does not always work. It requires in
future a new numerical method or a more radical reformulation such as
introducing a channel-dependent cutoff \cite{WAG} to resolve the problem.
Because the aim of the present evaluation is to compare characters of
different baryon-baryon interactions and not to do exact structure
calculations on the basis of the low-momentum equivalent interaction,
we present the results with the oscillatory behavior in case it appears.

$G$-matrix calculations for hyperons in symmetric nuclear matter are
carried out, using the continuous prescription for intermediate spectra.
Hyperon s.p. potentials are determined self-consistently. Details are
reported in Ref. \cite{KF}. In calculating the hyperon-nucleon $G$
matrices for chiral EFT, we use the nucleon s.p. potential obtained by
fss2 to focus on the properties of the hyperon-nucleon interactions.

\section{Calculated results}
We calculate equivalent $\Lambda N$, $\Sigma N$ and $\Xi N$ matrix elements
in the low-momentum space with the cut-off value of $\Lambda=2.0$ fm$^{-1}$
for the $^1S_0$ and $^3S_1$ partial waves, starting from the Kyoto-Niigata
SU$_6$ quark-model potential fss2 \cite{FSN} and the chiral
EFT potential \cite{CEFT1,CEFT2}. This momentum scale should be
regarded as a representative one for which the potential
dependence of the description of high momentum components has
been shown \cite{BOG} to disappear in the case of the $NN$ interaction.
As explained in Ref. \cite{KOKF}, we use the energy-independent version of
the quark-model potential \cite{SUZ} that eliminates the energy dependence
originating the RGM treatment of the quark clusters. Note that the
short-range part of the baryon-baryon interaction in the quark model
is constructed by a RGM framework for nonrelativistic quark-clusters,
while that of the chiral EFT potential is influenced by the contact
terms determined by phenomenological fitting.

\begin{figure}
\epsfxsize=5cm
\epsfbox{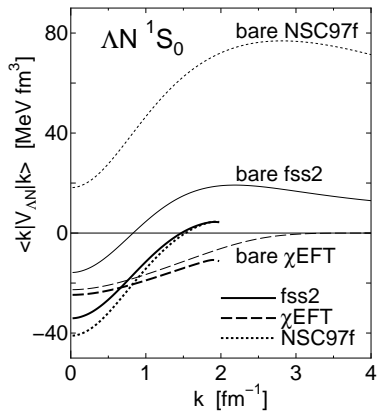}
\caption{Diagonal matrix elements of the equivalent interaction in the
low-momentum space with $\Lambda=2$ fm$^{-1}$ for the $\Lambda N$ $^1S_0$
channel, using the quark-model potential fss2 \cite{FSN}, the Nijmegen
potential NSC97 \cite{NSC}, and the chiral EFT potential ($\chi$EFT)
\cite{CEFT1} with a cut-off mass of 600 MeV.
Bare matrix elements are shown by thin curves. 
}
\end{figure}

\begin{figure}
\epsfxsize=5cm
\epsfbox{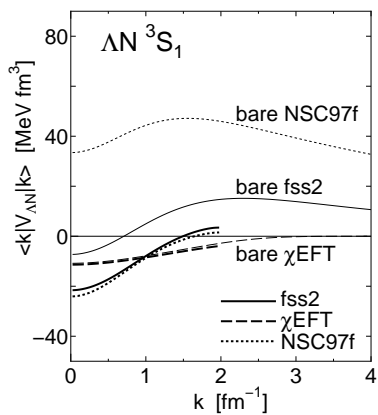}
\caption{Same as Fig. 1, but for the $\Lambda N$ $^3S_1$ channel.}
\end{figure}

\subsection{$\Lambda N$ interaction}

\begin{figure*}[th]
\epsfxsize=8.9cm
\epsfbox{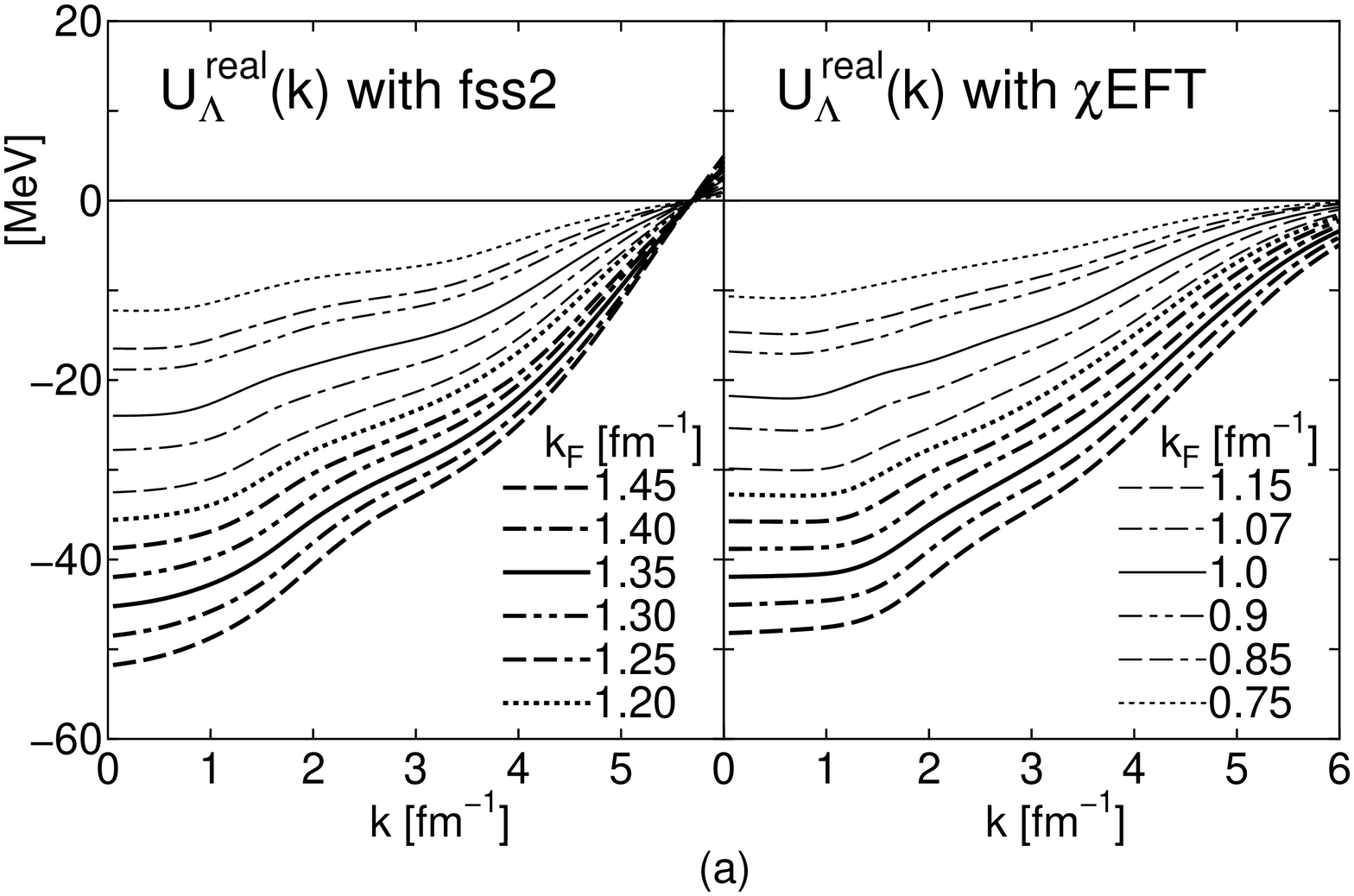}
\epsfxsize=8.9cm
\epsfbox{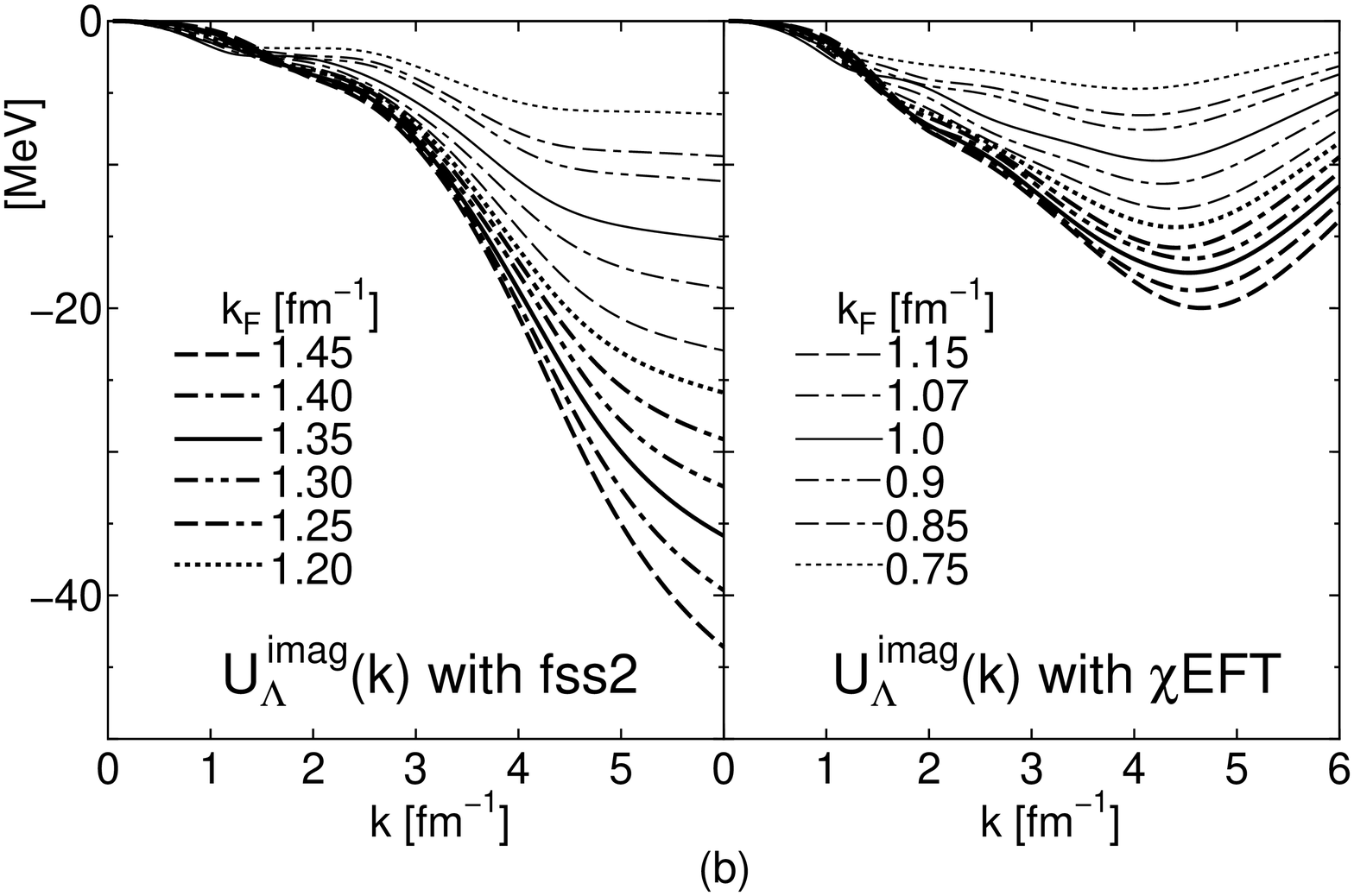}
\vspace*{-3mm}
\vspace*{-2mm}
\caption{Momentum dependence of $\Lambda$ s.p. potential in symmetric
nuclear matter at various Fermi momenta $k_F$: (a) real part and
(b) imaginary part. The calculations are in the lowest-order Brueckner
theory with the continuous prescription for the intermediate spectra.
The left panel shows the result of the quark-model potential
fss2 \cite{FSN}, and the right panel those of the potential of the
chiral EFT ($\chi$EFT) \cite{CEFT1} with a cut-off mass of 600 MeV.
}
\end{figure*}

Figures 1 and 2 show the low-momentum space diagonal matrix elements of
the equivalent $\Lambda N$ interaction in the $^1S_0$ and $^3S_1$ channels,
respectively, together with bare matrix elements. In this case we include
the equivalent interaction of the Nijmegen potential NSC97f \cite{NSC}, in
addition to the quark model potential fss2 \cite{FSN} and the chiral EFT
potential \cite{CEFT1} with a cut-off mass of 600 MeV.

As reported already in Ref. \cite{KOKF}, the NSC97f and the fss2 provide
very similar matrix elements in the low-momentum space, in spite of the
large difference in the short-range part as the bare matrix
elements indicate. On the other hand, the $k$-dependence of the chiral EFT
potential differs from other two potential, though the overall attractive
strength is of the same order. Because of the regularization with the cutoff
mass of $600$ MeV, the high-momentum component of the chiral EFT
potential is small and the equivalent interaction is not so much
different from the bare interaction in the low-momentum space.
The weak $k$-dependence suggests that the chiral EFT interaction is
more short-ranged than other two potentials in both
$^1S_0$ and $^3S_1$ channels. Note that in the $\Lambda N$ case,
a direct isovector $\pi$ exchange process is absent.
In the $^3S_1$ channel, a considerable amount of the attractive
contribution is expected from the $\Lambda N$-$\Sigma N$ coupling
through the $\pi$ exchange. In the cases of fss2 and NSC97f,
the attraction in a low-momentum space comes from this coupling
in a high-momentum space with the tensor component of
the $\pi$ exchange. In contrast, the small difference between bare
and low-momentum space matrix elements in the case of chiral EFT
implies that the coupling effect in a high-momentum
space is incorporated in the parameter of the contact terms.

Diagonal matrix elements of the effective interaction in momentum space
determine baryon s.p. potentials in nuclear matter. Because properties of
the s.p. potential can be more directly inferred from experimental data,
it is useful to present the calculated $\Lambda$ s.p. potential from the
$\Lambda N$ interaction. We can consider the Hartree potential obtained
by the equivalent interaction in the low-momentum space. However,
we prefer to use the standard lowest-order Brueckner theory, in which
some important many-body effects are incorporated. The calculated
$\Lambda$ s.p. potential from the two bare potentials, fss2 and chiral EFT,
are shown in Fig. 3. The real part is very similar in its magnitude
and $k_F$ dependence. It is not easy to detect the difference of
the $k$ dependence observed in the equivalent interactions in the
$^1S_0$ and $^3S_1$ channels. The imaginary part of the s.p. potential
indicates the strength of the $\Lambda N$-$\Sigma N$ coupling.
At low momentum, the chiral EFT potential gives slightly larger imaginary
strength. The weaker imaginary potential from the chiral EFT than that
from fss2 at large $k$ region is due to the weak $\Lambda N$-$\Sigma N$
coupling inherent in the cutoff mass of 600 MeV. 
As noted above, the coupling effect at high-momentum region may be
renormalized into the parameter of the contact terms in the chiral EFT
potential and thus the explicit $\Lambda N$-$\Sigma N$ coupling
at the large momentum region is weak in this parametrization.

As a whole, three bare potentials, fss2, NSC97f, and chiral EFT, for
the $\Lambda N$ interaction provide similar description of the $\Lambda$
s.p. potential. In literature \cite{MKGS,NOG06,FU08} we find that
the energy of the hypertriton is well reproduced by three potentials: namely
$E_{^3_\Lambda H}=-2.30$, $-2.487$, and $-2.34$ MeV for NSC97f, fss2,
and chiral EFT, respectively, compared with the empirical value of
$-2.354\pm 0.050$ MeV. However, the difference observed in Figs. 1 and 2
for the $k$ dependence of the equivalent interaction is probably detectable
in some experimental observables.

Finally, it is worth to comment on an unresolved problem of the microscopic
understanding of the small $\Lambda$ s.p. spin-orbit potential.
Experimentally, it has been established \cite{HT06} that the spin-orbit
splitting of the $\Lambda$ s.p. levels in nuclei is very small.
It is helpful to consider the Scheerbaum factor $S_\Lambda$ \cite{SCHE}
calculated in nuclear matter to relate the strength of the
$\Lambda$-nucleus spin-orbit potential to the two-body $\Lambda N$
interaction. The $\Lambda$-nucleus s.p. potential is well simulated by
\begin{equation}
 U_{\Lambda}^{\ell s}(r)=-\frac{\pi}{2} S_\Lambda \frac{1}{r} 
 \frac{d\rho (r)}{dr}\mbox{\boldmath{$\ell$}}\cdot\mbox{\boldmath{$\sigma$}},
\end{equation}
where $\rho (r)$ is a nucleon density distribution.
The necessary value of $S_\Lambda$ to explain experimental data
is about $-3.2$ MeV$\cdot$fm$^5$. In contrast, three potentials
considered here gives $S_\Lambda = -15.4, -12.2$ and $+4.8$
MeV$\cdot$fm$^5$ for NSC97f, fss2, and chiral EFT, respectively, in
normal symmetric nuclear matter, namely $k_F=1.35$ fm$^{-1}$.
The quark model suggests that antisymmetric spin-orbit component of the
two-body spin-orbit interaction may be important to cancel the ordinary
spin-orbit interaction. However, this mechanism does not work
quantitatively in fss2. Other two potentials do not contain the
antisymmetric spin-orbit component. It is significant that the chiral EFT
potential predicts an opposite sign for the $\Lambda$ s.p. spin-orbit
potential in the present leading order construction.

\begin{figure*}[t]
\epsfxsize=8.9cm
\epsfbox{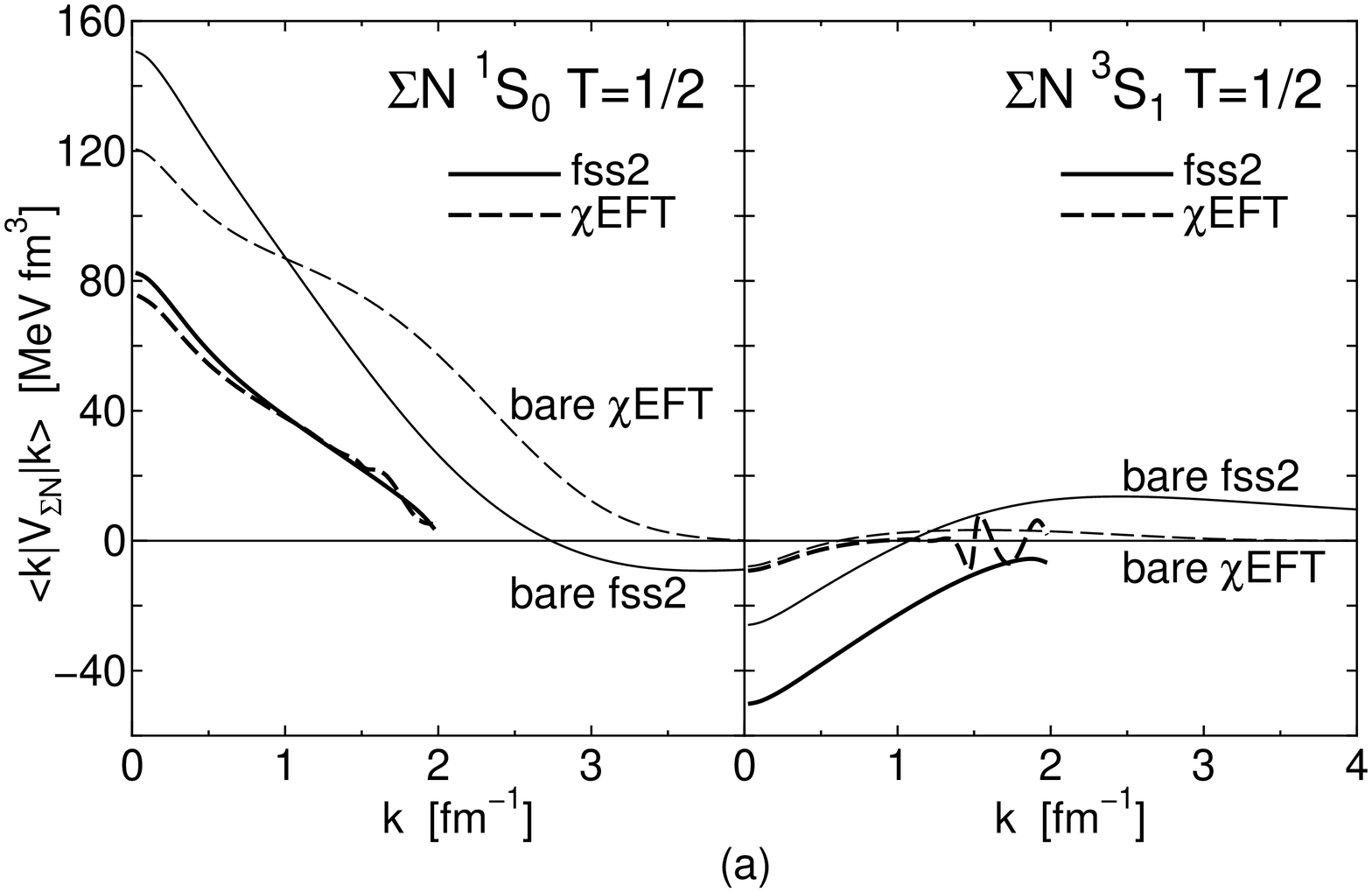}
\epsfxsize=8.9cm
\epsfbox{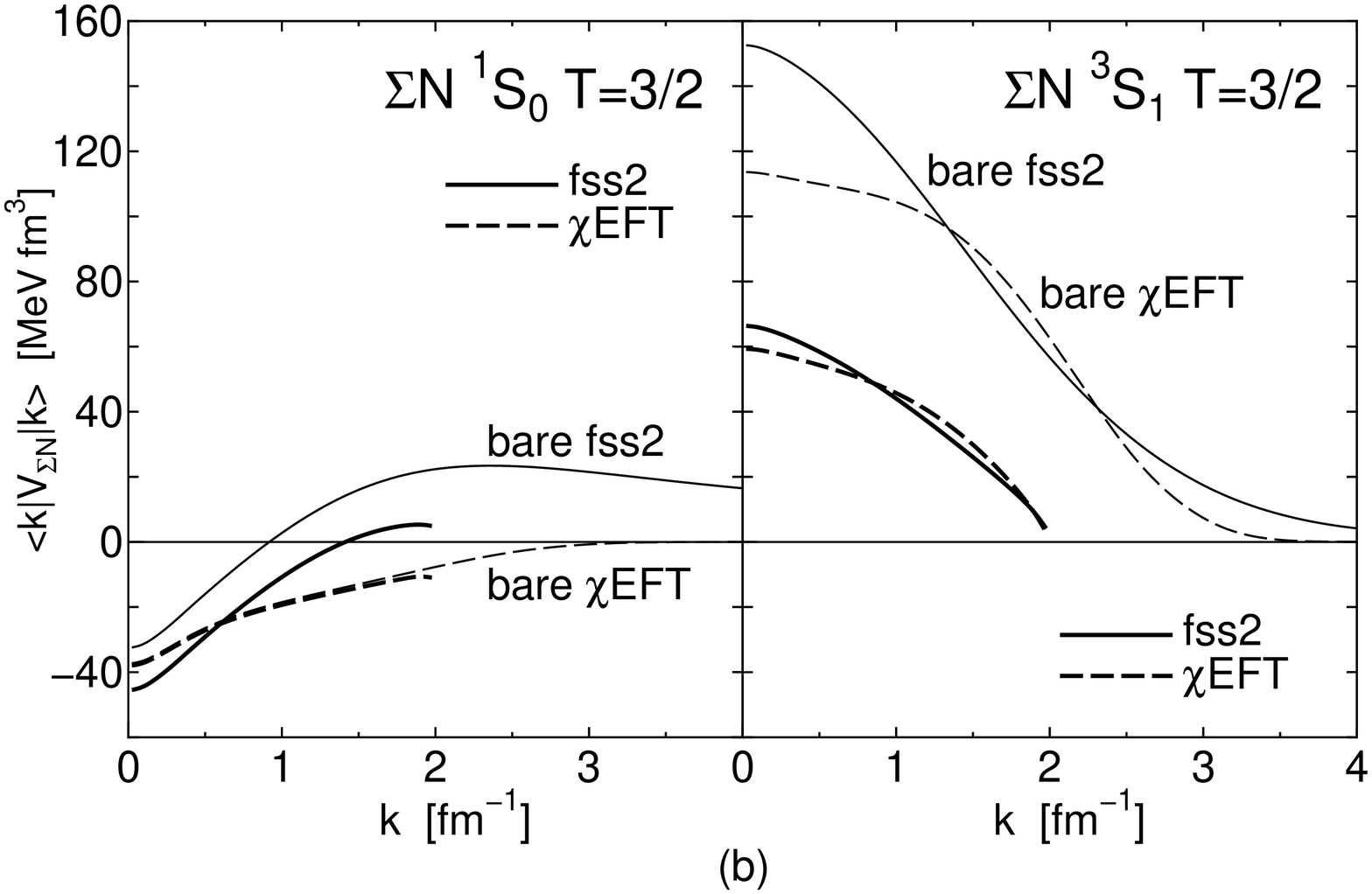}
\caption{Diagonal matrix elements of the equivalent interaction in the
low-momentum space with $\Lambda=2$ fm$^{-1}$ for the $\Sigma N$ $^1S_0$
and $\Sigma N$ $^3S_1$ channels, using fss2 \cite{FSN} and chiral EFT
($\chi$EFT) \cite{CEFT1}: (a) isospin $T=\frac{1}{2}$ and
(b) $T=\frac{3}{2}.$}
\end{figure*}

\begin{figure*}[ht]
\vspace*{5mm}
\epsfxsize=8.9cm
\epsfbox{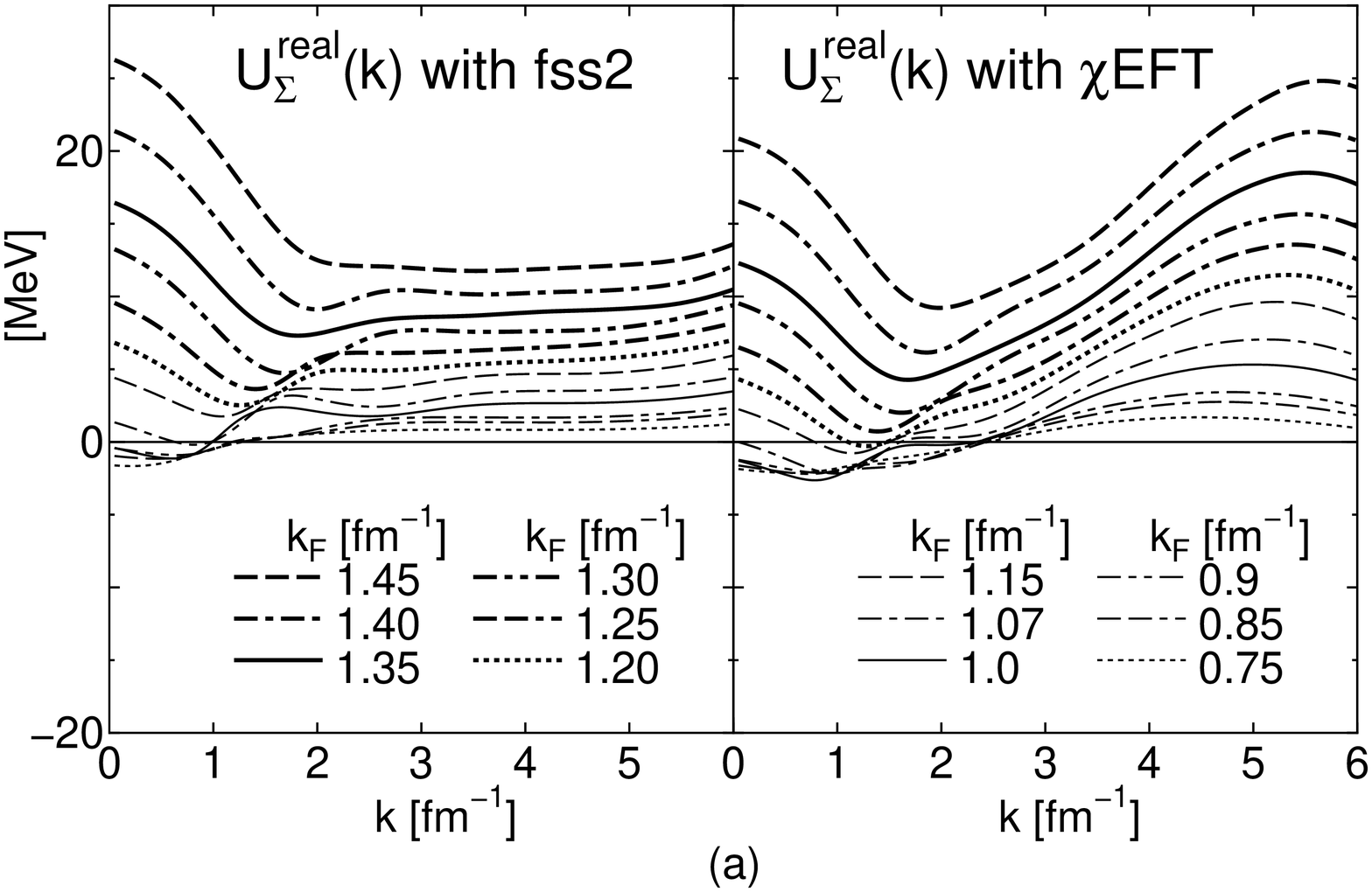}
\epsfxsize=8.9cm
\epsfbox{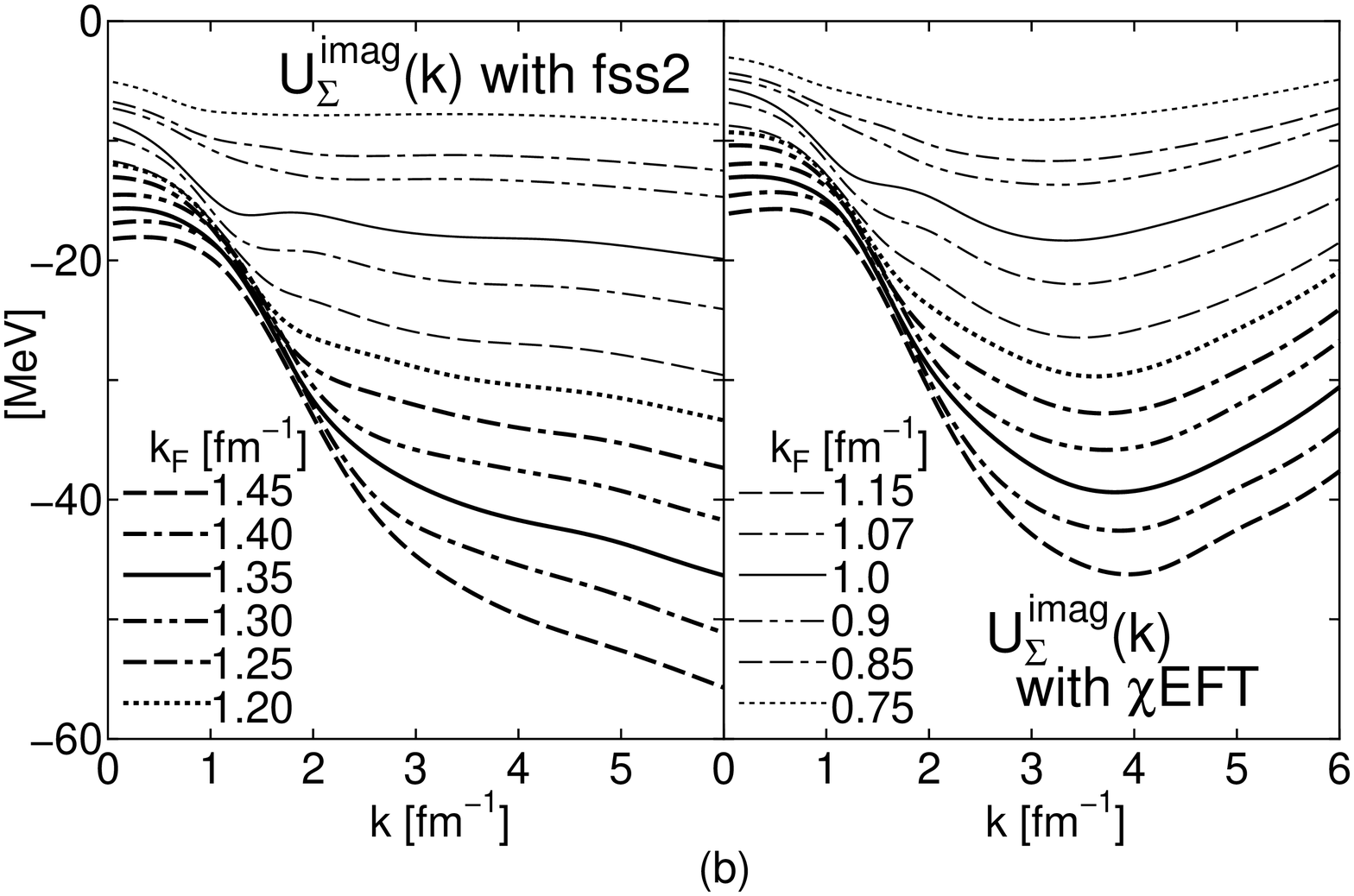}
\vspace*{-3mm}
\vspace*{-2mm}
\caption{Same as Fig. 3, but for the $\Sigma$ s.p. potential.
}
\end{figure*}

\subsection{$\Sigma N$ interaction}
Figure 4 shows the low-momentum space diagonal matrix elements of the
equivalent $\Sigma N$ interaction in the $^1S_0$ and $^3S_1$ channels for
the isospin $T=\frac{1}{2}$ and $\frac{3}{2}$, respectively, together with
bare matrix elements for the quark-model potential fss2 \cite{FSN} and
the chiral EFT potential with the cutoff mass of 600 MeV.

It is notable that the equivalent interactions of the quark model fss2 and
the chiral EFT potential are very similar except for the $^3S_1$
$T=\frac{1}{2}$ channel. It is known that the $\Sigma N$ $^1S_0$
$T=\frac{3}{2}$ state consists of the same $(2,2)$ flavor SU$_3$ symmetric
component of the Elliott notation $(\lambda, \mu)$ as the $NN$ $^1S_0$
state. Thus, this channel is expected to hold a rather strong attraction.
This character is manifested in the $J^\pi = 0^+$ $^4_\Sigma$He bound state
seen in the $^4$He$(K^-,\pi^-)$ reactions \cite{HAY,NAG}. The chiral EFT
potential also also has this attraction, although the $k$ dependence is
gentle as in the $\Lambda N$ equivalent interactions. 

The quark model picture has been known from the earlier studies
\cite{OSY,FU96a} to give a definite prediction that the $\Sigma N$
$^3S_1$ $T=3/2$ state should be strongly repulsive due to the quark Pauli
effect, which has no explicit counterpart in the OBEP parametrization.
The repulsive character persists in the low-momentum space. Owing to the
spin and isospin weight factors, this $^3S_1$ $T=3/2$ state dominantly
contributes to the $\Sigma$ s.p. potential in the nuclear medium,
as will be explicitly shown below in the calculated $\Sigma$ s.p.
potential. Analyses \cite{NOUM,HH,MK} of the $(\pi^-,K^+)$ $\Sigma$
formation inclusive spectra \cite{NOUM} have supported the overall
repulsive nature of the $\Sigma$-nucleus potential. Note that the
actual calculation \cite{KF09} in finite nuclei shows that the we
obtain weak attractive potential at the surface region of nuclei that
is necessary to account for the energy shift of $\Sigma^-$ atomic levels.

It is interesting to see that the matrix elements in the $^3S_1$
$T=\frac{3}{2}$ channel are similar for the fss2 and the chiral EFT.
While the repulsive character is dictated by the quark Pauli effect
in the fss2, the parameter of the contact term determined
phenomenologically is responsible for this repulsion  in chiral EFT. 

Calculated $\Sigma$ s.p. potentials in symmetric nuclear matter are
shown in Fig. 5. Two potentials predict very similar patterns for the
real part both in the $k$ dependence and in the $k_F$ dependence.
The size of the imaginary strength is also seen to be resembling except
for the region beyond $k \sim 4$ fm$^{-1}$.

\begin{figure*}[t]
\epsfxsize=8.9cm
\epsfbox{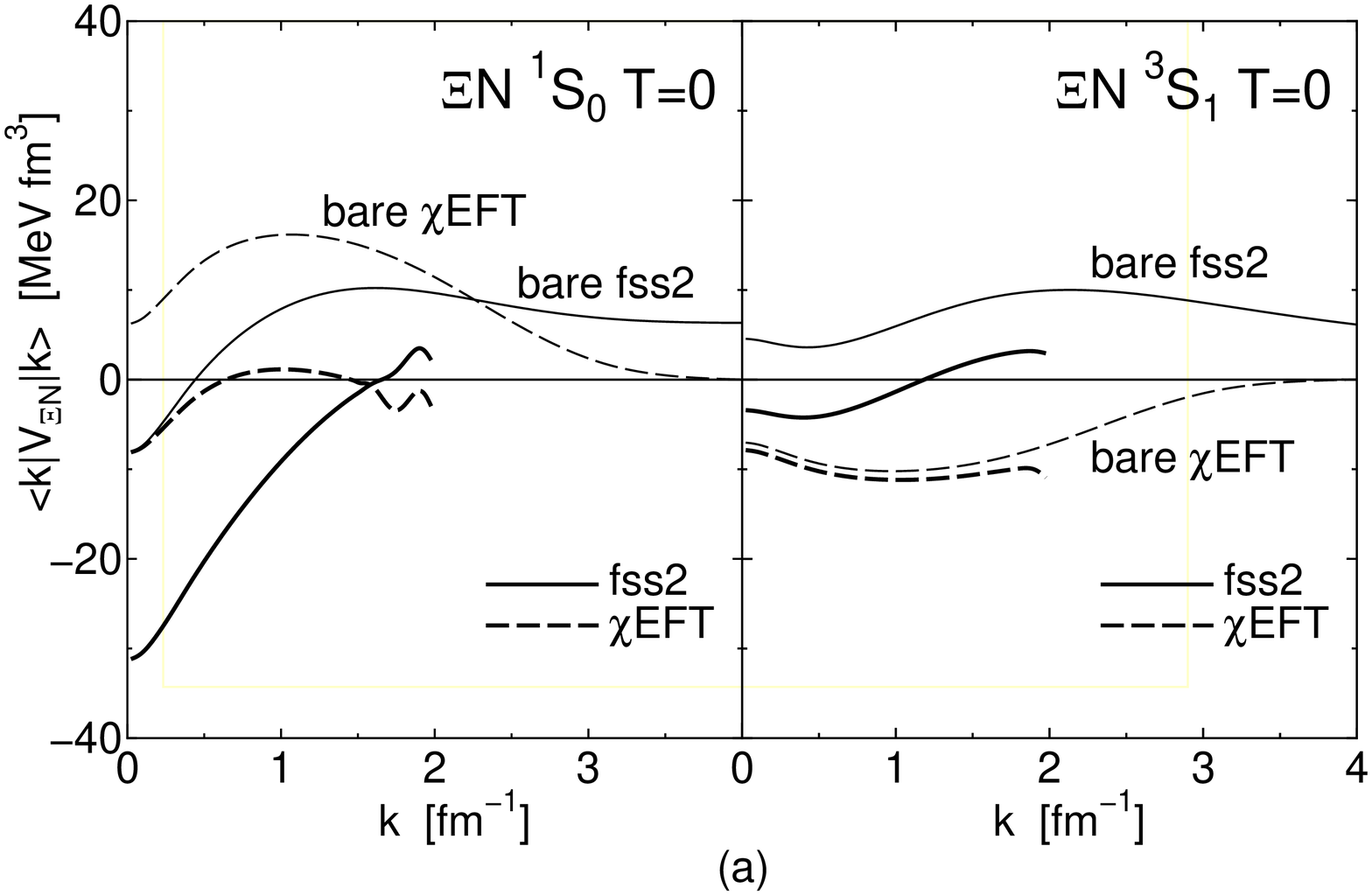}
\epsfxsize=8.9cm
\epsfbox{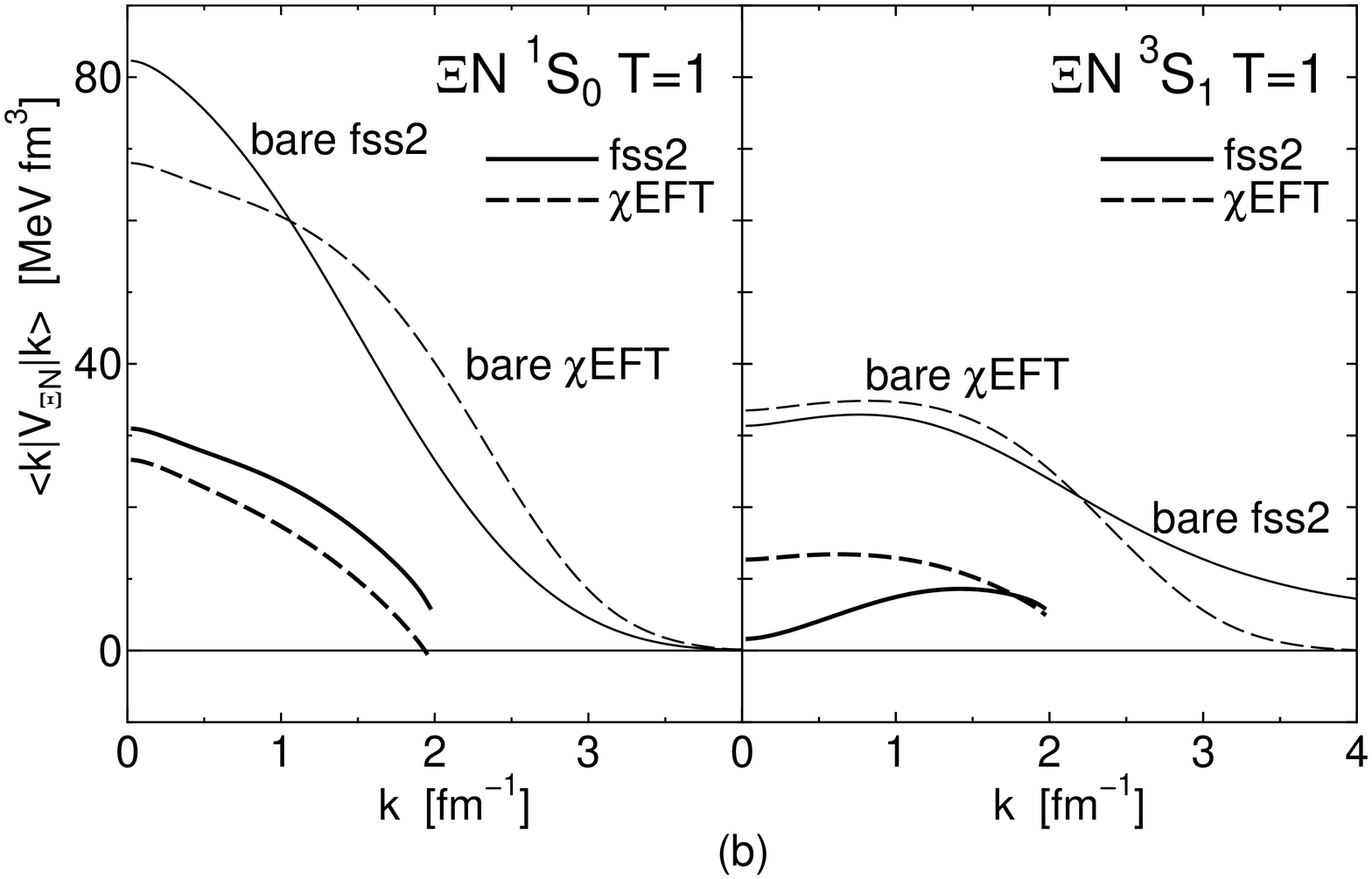}
\caption{Same as Fig. 4, but for the $\Xi N$ $^1S_0$ and $\Xi N$ $^3S_1$
channels: (a) isospin $T=0$ and (b) isospin $T=1$. The chiral EFT
($\chi$EFT) potential is from \cite{CEFT2}.
}
\end{figure*}

\begin{figure*}[ht]
\vspace*{5mm}
\epsfxsize=8.9cm
\epsfbox{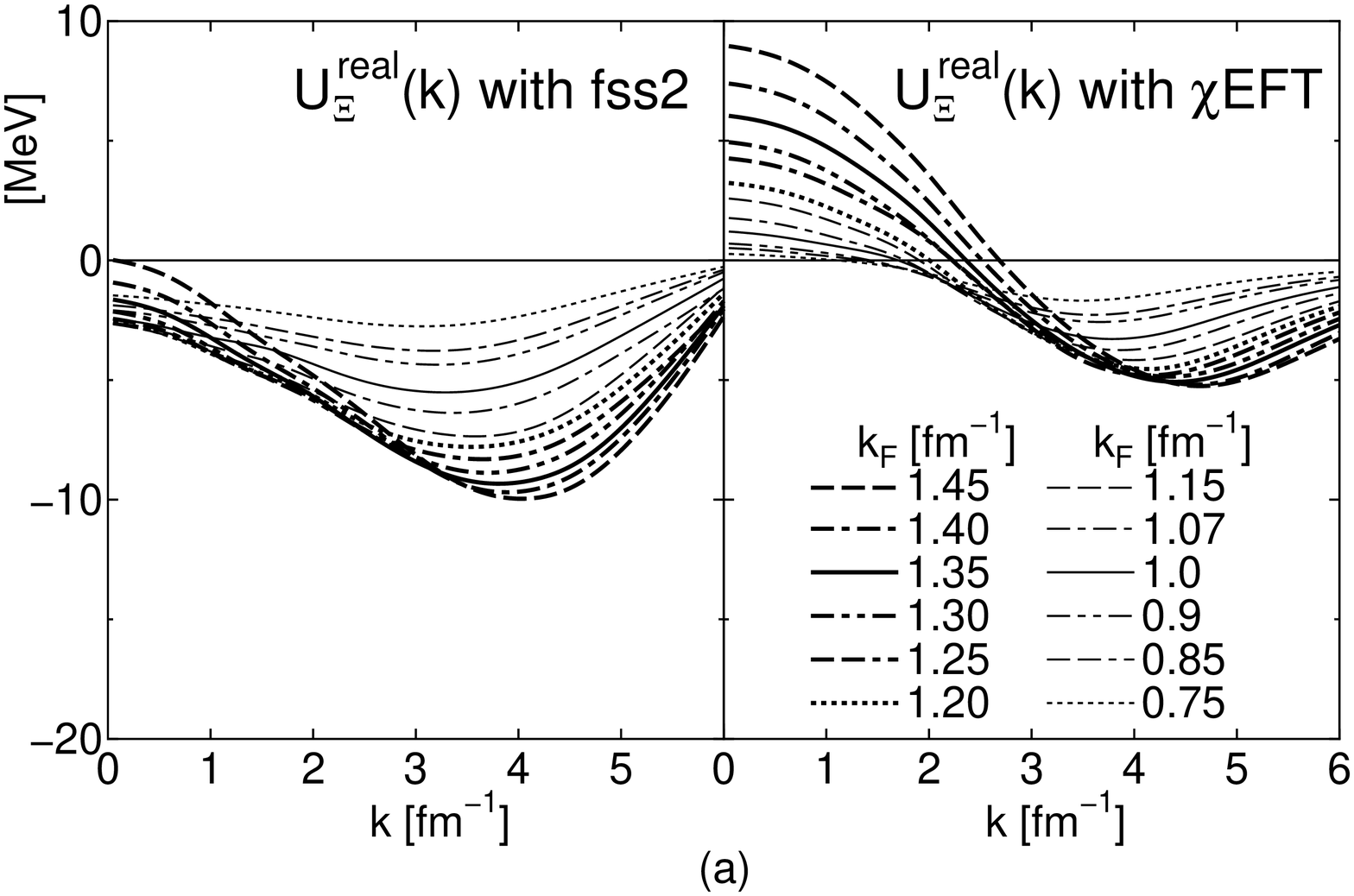}
\epsfxsize=8.9cm
\epsfbox{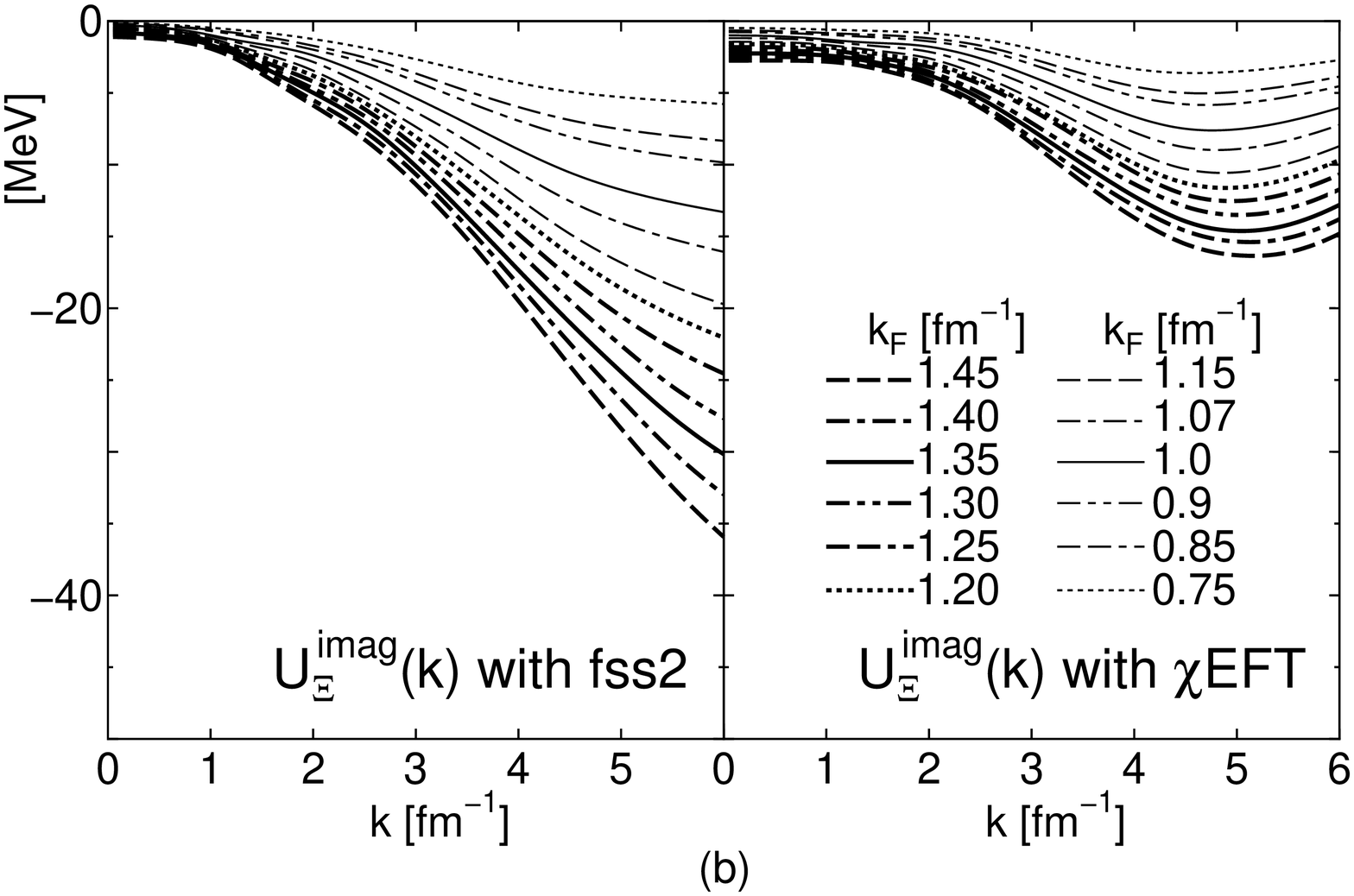}
\vspace*{-3mm}
\vspace*{-2mm}
\caption{Same as Fig. 3, but for the $\Xi$ s.p. potential. The chiral EFT
($\chi$EFT) potential is from \cite{CEFT2}.
}
\end{figure*}

\subsection{$\Xi N$ interaction}
Figure 4 shows the low-momentum space equivalent $\Xi N$ interaction
in the $^1S_0$ and $^3S_1$ channels for the isospin $T=\frac{1}{2}$ and
$\frac{3}{2}$, respectively, together with the bare matrix elements
up to $k=4$ fm$^{-1}$. In the $T=1$ channel, two potentials have similar
repulsive characters both in the bare and equivalent interactions.

The quark-model potential provides fair attraction in the $T=0$ $^1S_0$
channel. The most part of this attraction comes from
the $\Xi N$-$\Lambda \Lambda$-$\Sigma\Sigma$ coupling. This can be
checked by observing that if we switch off the baryon-channel coupling,
the matrix elements are close to those of the bare interaction.
In such a situation, it is important to consider the effect of the
baryon-channel coupling in the $P$ space to obtain more physically
meaningful information. The situation is the same in chiral EFT,
though the resulting attraction is very small in magnitude.
Note that in the chiral EFT theory an additional parameter has to be
introduced in the $^1S_0$ channel when extending to the $S=-2$ sector
from the $S=-1$ sector.

The $^3S_1$ $T=0$ state is classified to the pure $(11)_a$ state
in the flavor SU$_3$ symmetry and no baryon-channel coupling appears
in this state. The quark model \cite{FSN} predicts that the bare
$\Xi N$ interaction is already weak. Figure 11 shows that the
low-momentum equivalent $\Xi N$ interaction in this partial wave
becomes slightly attractive.

The quark-model potential fss2 \cite{FSN} predicts that the $\Xi N$
interactions in $^3S_1$ channels are weak. For the estimation of
the $\Xi$-nucleus s.p. potential in nuclear medium, we expect an
attractive contribution from the $^1S_0$ $T=0$ state but a repulsive
contribution from the $^1S_0$ $T=1$ state. Higher partial waves can
influence the sign of the $\Xi$-nucleus s.p. potential, although
it is unlikely that they bring about sizable net attractive or
repulsive contributions. The calculated $\Xi$ s.p. potentials
in symmetric nuclear matter in the LOBT are shown in Fig. 7.
The potential from the fss2 is seen to be weak. The tendency that
the attractive strength is largest at $k= 3\sim 4$ is owing to the
net contribution of $\Xi N$ $p$ waves.

Although the potential is attractive in infinite matter, the calculation
in finite nuclei \cite{KF09} shows that the $\Xi$ s.p. potential is
weakly attractive at the nuclear surface, and oscillates around zero
inside the nucleus. Judging from Fig. 7, a more repulsive $\Xi$ s.p.
potential in finite nuclei is expected from chiral EFT. The presently
available $(K^-,K^+)$ spectrum \cite{KH00} at the $\Xi^-$ production
threshold region is shown in Ref. \cite{KH09} to be consistent with
the weakly repulsive $\Xi$ potential. The prediction is to be
confronted with an experimental data with better accuracy obtained
soon from J-PARC \cite{JPC}.

\section{Summary}
We have compared two descriptions of the hyperon-nucleon
interactions, the Kyoto-Niigata quark-model potential fss2 \cite{FSN}
and the chiral EFT potential \cite{CEFT1,CEFT2}, by calculating
low-momentum space equivalent interactions and hyperon s.p. potentials
in the LOBT in symmetric nuclear matter obtained from these bare
potentials. The purpose is to elucidate the similarity and the difference
in the $\Lambda N$, $\Sigma N$ and $\Xi N$ interactions between the quark
model and the chiral EFT theory. The former model is based on a
resonating-group method for two constituent-quark clusters with an
effective gluonic interaction and long-ranged one-boson exchanges
between quarks. The energy dependence inherent in the RGM treatment
is eliminated by the method in Ref. \cite{SUZ}. The latter
parametrization uses pseudoscalar-meson exchanges and flavor SU$_3$
invariant contact terms, regularized by a cut-off mass of around 600 MeV.
Parameters of the contact terms, 5 in the $S=-1$ sector and an additional
one parameter in the $S=-2$ sector, are determined by fitting to
available experimental data. Because of the difference in the description
for the short-range part, it is worthwhile to compare two potentials.

In the previous paper \cite{KOKF}, we showed that the $\Lambda N$
equivalent interaction in the low-momentum space is almost identical
for the quark model fss2 and the Nijmegen OBEP model NSC97f \cite{NSC}.
In this paper, we have found that the leading order chiral EFT
interaction gives matrix elements of the equivalent interaction which
have different $k$ dependence from the fss2. This difference is not
visible in the $\Lambda$ s.p. potential in the nuclear medium,
although it is probably detectable in some observables in future
experiments. Note that there is an unresolved problem of describing
very small spin-orbit splitting of the $\Lambda$ hyperon in nuclei.
$G$-matrix calculations, in which effects of the $\Lambda N$-$\Sigma N$
coupling in the nuclear medium are taken care of, show that fss2 does
not provide a small $\Lambda$-nucleus spin-orbit potential necessary
to account for the empirical data, in spite of the tendency of the
cancellation of the ordinary and antisymmetric spin-orbit components.
On the other hand, the chiral EFT potential having no antisymmetric
spin-orbit component predicts a small but opposite sign for the
spin-orbit potential. It is interesting if the effects of the next
leading prder corrections is revealed in the future analysis.

As for the $\Sigma N$ interactions, the quark model fss2 and the chiral
EFT potential mostly give similar matrix elements of the equivalent
interactions in a low-momentum space, except for the $^3S_1$ $T=\frac{1}{2}$
state. It is interesting to see that the repulsion in the $^3S_1$
$T=\frac{3}{2}$ state predicted by the quark model because of the quark
Pauli effect is reproduced as well in the chiral EFT parametrization.
This similarity reflects in that calculated $\Sigma$ s.p. potentials
in symmetric nuclear matter also resemble in their magnitude, momentum
dependence, and $k_F$ dependence.
The prediction of the repulsive potential for the $\Sigma$ hyperon
embedded in the nuclear medium is not common among several baryon-baryon
interaction parametrizations. Therefore checks by forthcoming experiments,
for example in the J-PARC project \cite{JPC}, will be very important for
understanding the $\Sigma N$ interaction.

The resemblance of the two interactions holds also in the $\Xi N$
interaction. The fss2 interaction provides weakly attractive s.p.
potentials in symmetric nuclear matter. The chiral EFT interaction
tends to give slightly more repulsive s.p. potentials because of the
lack of attraction in the $^1S_0$ $T=0$ channel. The microscopic
calculation \cite{KF09} of the $\Xi$-nucleus potential in finite
nuclei shows that the fss2 predicts an almost zero potential.
At the surface region, the potential is weakly attractive and inside
the nucleus the potential fluctuate around zero. Such weak $\Xi$-nucleus
potential is shown in Ref. \cite{KH09} to be able to account for the
existing $(K^-,K^+)$ spectrum at the threshold region \cite{KH00}.
This experimental data is based on the small number of counts and
thus may not be accurate enough to conclude the strength of the
$\Xi$-nucleus potential. We expect new $\Xi$-production spectrum data
with better accuracy from the J-PARC experiments \cite{JPC}, which
will provide important information on the baryon-baryon interaction
in the $S=-2$ sector.
\\

\acknowledgments
The author is grateful to Y. Fujiwara, H. Polinder, and J. Haidenbauer
for providing him computational codes of their baryon-baryon interactions.

\end{document}